\documentclass[twocolumn,superscriptaddress,floatfix,preprintnumbers, nofootinbib,hyperref]{revtex4-2} 
\pdfoutput=1
\usepackage[colorlinks=true,breaklinks=true]{hyperref}
\usepackage[normalem]{ulem}
\usepackage[utf8]{inputenc}
\hypersetup{allcolors=[rgb]{0.0 0.0 0.6},linkcolor=[rgb]{0.75 0.05 0.05}}
\usepackage{amsmath,amssymb}
\usepackage{epsfig}  
\usepackage{graphicx}   
\usepackage{slashed}       
\usepackage{tikz}
\usepackage{tikz-feynman}
\usepackage{url}
\usepackage{color}
\usepackage{multirow}
\usepackage{comment}
\usepackage{amssymb}

\DeclareMathOperator{\GeV}{GeV}
\DeclareMathOperator{\eV}{eV}

\DeclareMathOperator{\keV}{keV}
\DeclareMathOperator{\MeV}{MeV}

\DeclareMathOperator{\TeV}{TeV}

\newcommand{\beq}{\begin{equation}}
\newcommand{\eeq}{\end{equation}}

\allowdisplaybreaks

\bibpunct{[}{]}{,}{n}{}{,}

\pgfarrowsdeclare{X}{X}
{
  \pgfutil@tempdima=0.3pt%
  \advance\pgfutil@tempdima by.25\pgflinewidth%
  \pgfutil@tempdimb=5.5\pgfutil@tempdima\advance\pgfutil@tempdimb by.5\pgflinewidth%
  \pgfarrowsleftextend{+-\pgfutil@tempdimb}
  \pgfarrowsrightextend{0pt}
}
{
  \pgfutil@tempdima=0.3pt%
  \advance\pgfutil@tempdima by.25\pgflinewidth%
  \pgfsetdash{}{+0pt}
  \pgfsetroundcap
  \pgfsetmiterjoin
  \pgfpathmoveto{\pgfqpoint{-5.5\pgfutil@tempdima}{-6\pgfutil@tempdima}}
  \pgfpathlineto{\pgfqpoint{5.5\pgfutil@tempdima}{6\pgfutil@tempdima}}
  \pgfpathmoveto{\pgfqpoint{-5.5\pgfutil@tempdima}{6\pgfutil@tempdima}}
  \pgfpathlineto{\pgfqpoint{5.5\pgfutil@tempdima}{-6\pgfutil@tempdima}}
  \pgfusepathqstroke 
}
\makeatother

\begin{document}
\title{Glueball Axion-Like Particles}

\author{Pierluca Carenza}\email{pierluca.carenza@fysik.su.se}
\affiliation{The Oskar Klein Centre, Department of Physics, Stockholm University, Stockholm 106 91, Sweden}

\author{Roman Pasechnik}\email{roman.pasechnik@fysik.lu.se}
\affiliation{
 Department of Physics, Lund University\\
 SE-223 62 Lund, Sweden
}

\author{Zhi-Wei Wang}\email{zhiwei.wang@uestc.edu.cn}
\affiliation{School of Physics, The University of Electronic Science and Technology of China,\\
 88 Tian-run Road, Chengdu, China}

\smallskip

\begin{abstract}
Dark Yang-Mills sectors that confine to form stable composite states, known as glueballs, have been traditionally proposed as a potential explanation for cosmological Dark Matter (DM). Earlier studies have established viability of the lightest scalar glueball as a possible DM candidate. In this work, we explore a whole class of effective composite sectors in the confined Yang-Mills regime featuring an additional pseudoscalar glueball state. We also investigate the role of effective interactions of the dark glueball sector with the visible sectors via higher-dimensional operators primarily focusing on dimension-8 couplings of glueballs to photons and gluons. We stress the remarkable similarities between the phenomenology of such glueball effective theories and  Standard Model extensions featuring Axion-Like Particles (ALPs). Hence, one deals with a new class of composite Glueball ALPs (or GALPs) coupled to photons and/or nucleons in a wide mass range, from sub-eV to the Planck scale, yielding viable DM candidates that can be probed by astrophysical and cosmological observations. 
\end{abstract}

\maketitle

\section{Introduction}

Dark Matter (DM) in the form of glueball composite states arising from confinement of dark Yang-Mills (YM) sectors are widely studies in the literature~\cite{Carenza:2022pjd,Carenza:2023eua,Carlson:1992fn,Faraggi:2000pv,Feng:2011ik,Boddy:2014yra,Soni:2016gzf,Kribs:2016cew,Acharya:2017szw,Dienes:2016vei,Soni:2016yes, Soni:2017nlm,Draper:2018tmh,Halverson:2018olu, Forestell:2017wov, Forestell:2016qhc,McKeen:2024trt,Biondini:2024cpf,Das:2018ons}. Such sectors cool down with the Universe expansion and, at low temperatures, feature a first-order confinement-deconfinement phase transition occurring at a critical temperature $T_c$~\cite{Panero:2009tv,Halverson:2020xpg,Huang:2020crf,Reichert:2021cvs,Kang:2021epo}. Characterizing the phase transition is challenging and requires a delicate interplay between thermal field theory and non-perturbative phenomena.

In previous studies~\cite{Carenza:2022pjd,Carenza:2023eua}, the authors have computed the scalar glueball relic density emerging from $SU(N)$ gauge theories, for the number of colors $N=\{3,4,5\}$. The used formalism employed a low-energy effective field theory (EFT) of gluon-glueball dynamics in the framework of Polyakov loop model~\cite{Sannino:2002wb}, proving that the lightest scalar glueball can be a suitable DM candidate. More recently, in Ref.~\cite{Carenza:2024qaq} the authors have introduced another relevant degree of freedom in the low-energy description of the YM theory: a pseudoscalar glueball state that has been subject of intense investigation in the context of QCD hadron physics, see e.g.~Refs.~\cite{Rosenzweig:1981cu,Eshraim:2012jv,Cheng:2008ss,Li:2007ky,Masoni:2006rz,Kochelev:2005vd,Gabadadze:1997zc,Chowdhury:2014mra}. In this work, we extend upon Ref.~\cite{Carenza:2024qaq} by providing more details on the derivation of the effective potential determining the dynamics of both the lightest scalar and pseudoscalar glueball states. Moreover, we discuss possible interactions of these states with the visible sector, expanding the discussion in Ref.~\cite{Carenza:2024qaq} to show the existing constraints on glueballs coupled to photons and nucleons, even in the case when they do not constitute the entirety of DM.

The GALP interactions considered in our analysis stem from EFT interaction operators between the dark YM sector and the Standard Model (SM) particles, induced by a heavy fermionic mediator at the microscopic level. Specifically, we focus on glueball interactions with massless SM gauge bosons -- photons and gluons -- via effective dimension-8 operators acting as a ``gauge portal'' between the dark YM and visible SM sectors~\cite{Faraggi:2000pv}, leaving the study of ultraviolet-complete models involving dark fermions and massive SM vector bosons to a future work. Once the dark YM sector confines, such operators induce effective interactions of glueballs with photons and gluons. We notice that these interactions are very similar to the ones arising in Axion-Like-Particle (ALP) models~\cite{Svrcek:2006yi,Cicoli:2012sz,Halverson:2019kna}. Thus, the phenomenology of such weakly-coupled glueballs is similar to the one of ALPs enabling us to refer to such states as Glueball ALPs (or GALPs in what follows). Just like in the case of ALPs~\cite{Preskill:1982cy,Abbott:1982af,Dine:1982ah}, the GALPs scenario provides viable DM candidates and interesting phenomenology that will be discussed below. Besides, we will point out most relevant differences between ALPs and GALPs, in terms of their properties, interactions and phenomenological implications.

This work is organized as follows. In Sec.~\ref{sec:efflag} we derive from the first principles the EFT Lagrangian involving the scalar and pseudoscalar glueballs. In Sec.~\ref{sec:relic} we discuss how, based on the previous studies of Refs.~\cite{Carenza:2022pjd,Carenza:2023eua}, it is possible to estimate the relic glueball DM abundance. In Sec.~\ref{sec:GALP} we explore possible interactions between the glueballs and the SM particles, in particular, with photons and gluons. We introduce the concept of GALPs whose phenomenology is similar to that of ALPs and discuss some of the basic constraints on their parameter space. In Sec.~\ref{sec:relation} we introduce the mass-coupling relation for GALPs, characterizing this model in analogy with axion models. In Sec.~\ref{sec:differences} we discuss the most important differences between GALPs and axions.
Finally, in Sec.~\ref{sec:conclusions} we summarize our findings and conclude. An appendix then follows, with details on the glueball effective potential.

\section{Dark gluons and glueballs} 
\label{sec:efflag}

In the following, we will consider a dark gauge $SU(N)$ sector, with $N\ge 3$ colors. The physics of the associated ``dark gluon" fields $A_{\mu}^{a}$ is described by the Lagrangian,
\begin{equation}
\begin{split}
    \mathcal{L}_{\rm SU(N)}&=-\frac{1}{4}G_{\mu\nu}^{a}G^{\mu\nu a}+\frac{\theta }{4}G_{\mu\nu}^{a}\tilde{G}^{\mu\nu a}\,,\\
    G_{\mu\nu}^{a} &= \partial_\mu A_{\nu}^{a} - \partial_\nu A_{\mu}^{a} + g f^{abc} A_{\mu}^{b} A_{\nu}^{c} \,, \\ 
    a,b,c &= 1,\dots,N^2-1 \,,
    \label{eq:lagrangian}
\end{split}
\end{equation}
in terms of $g=g(\mu)$ and $f^{abc}$ being the gauge coupling evolving with the scale $\mu$ and the gauge structure constants, respectively, while the field strength tensor and its dual tensor are denoted as $G_{\mu\nu}^{a}$ and $\tilde{G}_{\mu\nu}^{a}=\frac{1}{2}\epsilon_{\mu\nu\alpha\beta}G^{\alpha\beta a}$, respectively. We are interested in the case with Charge-Parity (CP) not being \emph{a priori} a symmetry of the dark sector as implied by the term proportional to the $\theta$-parameter.

The theory given by Eq.~\eqref{eq:lagrangian} is asymptotically free in the ultraviolet regime \cite{Politzer:1973fx,Gross:1973ju,Gross:1974cs}, while in the infrared limit it exhibits confining behavior (i.e.~gluons are confined into glueballs) when the energy scale $\mu$ is below the confinement scale $\Lambda$ \cite{Nambu:1974zg,Polyakov:1975rs,tHooft:1974kcl}. On the other hand, the $SU(N)$ gauge theory of Eq.~\eqref{eq:lagrangian} can also be generalized to other groups (see Refs.~\cite{Yamada:2022imq,Yamada:2022aax,Bruno:2024dha}). In analogy to the gauge coupling, principle, $\theta$ may run with $\mu$~\cite{Reuter:1996be,Shifman:1990nz,Knizhnik:1984kn}, but for the sake of simplicity we ignore this possibility in what follows.

We adopt a minimal setup analogical to ordinary QCD and introduce in addition a heavy Dirac state denoted as $\Psi$ charged under dark $SU(N)$ and, possibly, under other SM gauge symmetries. In this work, we consider $\Psi$ being coupled to photons and gluons, thus, playing a role of a ``portal'' between the SM and dark sectors. While its Yukawa interaction with the Higgs boson may also be introduced, we leave this option for future studies. To ensure highly suppressed (feeble) visible-to-dark interactions at low energies, such a Dirac state can be assumed very heavy, with mass well above the $SU(N)$ confinement scale $\Lambda$, i.e.~$M_\Psi \gg \Lambda$, such that its impact on confining dynamics can be safely ignored. Integrating out such a massive portal, one naturally arrives at the set of higher-dimensional operators mediating effective interactions between the visible sectors and dark gluons in low-energy limit of such a theory. The lightest composite spectrum of the dark sector relevant at low energies would be made of dark glueballs which are the subject of analysis in this work.

Indeed, due to confinement a tower of color-neutral glueballs is generated at temperatures below the critical one for the confinement-deconfinement phase transition in the dark gauge sector. Such states are characterized by various composite $SU(N)$-invariant operators. Thermal dynamics of the lightest scalar glueballs, ${\cal H}\equiv 0^{++}$, in expanding Universe has been discussed recently by the authors in~\cite{Carenza:2022pjd,Carenza:2023eua} in effective $SU(N)$ ($N=3,4,5$) theories. The dynamics was captured by an EFT approach~\cite{Sannino:2002wb} describing a composite scalar glueball sector interacting with the high-temperature gluon gas represented in terms of the Polyakov loop. In this work, following Ref.~\cite{Carenza:2024qaq}, we consider also the pseudoscalar glueball denoted as ${\cal A}\equiv 0^{-+}$, with properties analogical to those of conventional ALPs, thus, suggesting their potentially relevant role as a DM candidate.

Explicitly, the following lowest-dimension gauge-invariant operators can give rise to the $0^{++}$ and $0^{-+}$ glueball states,
\begin{equation}
   {\cal A}\, {\cal H}^{3} \equiv G^{a}_{\mu\nu}\tilde{G}^{\mu\nu a}\,, \qquad 
   {\cal H}^{4} \equiv -\frac{\beta(g)}{2g} G^{a}_{\mu\nu}G^{\mu\nu a}
  \,.
  \label{GG-to-glue}
\end{equation}
The effective interactions in the low-energy limit of the theory are driven by the EFT Lagrangian~\cite{Rosenzweig:1979ay,Schechter:1980ak}
\begin{equation}
    \mathcal{L}_{\rm eff}=\frac{1}{2}\partial_{\mu}{\cal H}\partial^{\mu}{\cal H}+\frac{1}{2}\partial_{\mu}{\cal A}\partial^{\mu}{\cal A}-V_{\rm eff}({\cal H},{\cal A})\,,
    \label{eq:efflagV}
\end{equation}
in terms of the glueball potential $V_{\rm eff}$. Furthermore, the EFT in this form features the dilatonic current non-conservation property of the dark $SU(N)$ theory~\cite{Peskin:1995ev,Collins:1976yq,Rosenzweig:1979ay}
\begin{equation}
    \begin{split}
\partial_{\mu}D^{\mu}&=\frac{\beta(g)}{2g}G_{\mu\nu}^{a}G^{\mu\nu a}=-{\cal H}^4\,,
\label{eq:noncons}
    \end{split}
\end{equation}
where $\beta(g)$ is the $\beta$-function of the running $SU(N)$ gauge coupling $g(\mu)$. The property (\ref{eq:noncons}) holds in the non-perturbative regime, thus, constraining the form of the glueball effective potential $V_{\rm eff}$.

Dilatation transformations for a glueball field $\varphi \equiv \{{\cal H},{\cal A}\}$ read
\begin{equation}
    \varphi'(x')=\varphi'(\lambda^{-1}x)=\lambda\, \varphi(x) \,,
\end{equation}
and the corresponding dilatonic current $D^{\mu}$ can be found starting from Eq.~\eqref{eq:efflagV} as
\begin{equation} 
\partial_{\mu}D^{\mu} = -\Theta^{\mu}_{\mu}=4V_{\rm eff}-\frac{\partial V_{\rm eff}}{\partial {\cal H}}{\cal H}-\frac{\partial V_{\rm eff}}{\partial {\cal A}}{\cal A}\,,
\label{eq:dilaton}
\end{equation}
in terms of the energy-momentum tensor in canonical form $\Theta_{\mu\nu}$. Eqs.~\eqref{eq:noncons} and \eqref{eq:dilaton} enable one to obtain the constrained form of the glueball potential \cite{Carenza:2024qaq},
\begin{equation}
\begin{split}
    V_{\rm eff}&\simeq c_{0}{\cal H}^{4}\ln\left(\frac{{\cal H}}{\Lambda}\right)+\frac{\theta}{4}{\cal A}\,{\cal H}^{3}+{\cal H}^{4}f\left(\frac{{\cal A}}{{\cal H}}\right)+\\
&\quad+c_{1}{\cal H}^{4}+c_{2}{\cal H}^{2}{\cal A}^{2}+c_{3}{\cal A}^{4}+c_{4}{\cal H}\,{\cal A}^{3}\,,
\end{split}
    \label{eq:potential}
\end{equation}
where continuous function $f$ remains arbitrary. Due to a non-zero $\theta$-term on the fundamental gauge theory~\eqref{eq:lagrangian}, odd powers of ${\cal A}$, specifically those proportional to $\theta$ and $c_{4}$, are allowed in the potential~\eqref{eq:potential} giving rise to CP-violation in the glueball EFT. The form of arbitrary function $f$, as well as effective parameters $\theta$ and $c_{i}$, $i=0,\dots,4$ are expected to be constrained by the glueball phenomenology.

In order to gain a basic understanding of the glueball dynamics, it is instructive to consider small quantum fluctuations about the classical glueball vacuum defined by the vacuum expectation values (VEVs) $\varphi_0 \equiv \{\eta_{0}, a_{0}\}\sim\mathcal{O}(1)\Lambda$. Expanding the effective potential into series over glueball excitations $\eta\equiv {\cal H}-\eta_{0}$ and $a\equiv {\cal A}-a_{0}$ about the ground state, it is straightforward to deduce \cite{Carenza:2024qaq}
\begin{equation}
  V_{\rm eff} \simeq \frac{m_{\eta}^{2}}{2} \eta^{2} + \frac{m_{a}^{2}}{2}a^{2}+\sum_{i=0}^{\infty}\sum_{j=0}^{\infty}\frac{\lambda_{ij}}{\Lambda^{i+j-4}}\eta^{i} a^{j}\Bigg|_{i+j\ge3}\,.
  \label{eq:exp}
\end{equation}
Hence, we arrive at the glueball spectrum in terms of the physical mass eigenstates where $m_{\eta}$ and $m_{a}$ are their masses, while interactions below the confinement energy scale $\Lambda$ are effectively determined by dimensionless parameters $\lambda_{ij}$. It is important to note that $i + j > 4$ terms are suppressed by higher powers of $\Lambda$, and are therefore negligible at phenomenologically relevant energy scales.

\section{Glueball dark matter}
\label{sec:relic}

Many aspects of a DM sector composed of dark glueballs have been extensively studied so far~(see e.g.~Refs.~\cite{Carenza:2022pjd,Carenza:2023eua,Carlson:1992fn,Faraggi:2000pv,Feng:2011ik,Boddy:2014yra,Soni:2016gzf,Kribs:2016cew,Acharya:2017szw,Dienes:2016vei,Soni:2016yes, Soni:2017nlm,Draper:2018tmh,Halverson:2018olu, Forestell:2017wov, Forestell:2016qhc,Yamada:2023thl,McKeen:2024trt,Biondini:2024cpf} and references therein). In this work, we will discuss phenomenology of our two-glueball effective model in Eq.~\eqref{eq:exp} considering two possible scenarios of one- and two-component DM.

In the case of DM composed by stable glueballs of one type, one adopts the hypothesis that at temperatures $T\lesssim T_c$ an in-medium glueball decay mode was allowed such that heavier (e.g.~pseudoscalar $a$) glueballs decayed into lightest (e.g.~scalar $\eta$) ones. The resulting relic density of lightest glueballs can be estimated as~\cite{Carenza:2022pjd,Carenza:2023eua}
\begin{equation}
    \Omega_{\rm DM}h^{2}=0.12\,\zeta_{T}^{-3}\frac{\Lambda}{\Lambda_{0}}\,,\quad 
    100~\eV\lesssim\Lambda_{0}\lesssim 400~\eV \,.
\label{eq:lambda0}
\end{equation}
Its linear scaling with $\Lambda$ is at the origin of the dark glueball overabundance~\cite{Halverson:2016nfq}. To match the observed DM abundance, one typically assumes a suppressed dark-to-visible temperature ratio $\zeta_{T}^{-1}$. It is worth noting here that $\Lambda$ can take any values within the range~\cite{Carenza:2023eua}
\begin{equation}
20~\MeV \lesssim \Lambda \lesssim 10^{10}~\GeV\,,
\label{Lambda-const}
\end{equation}
where the lower limit ensures that DM self-interactions are not excessively strong, while the upper limit requires the dark confinement-deconfinement phase transition to happen after inflation such that DM does not get exponentially diluted.

Due to a strongly first-order confinement-deconfinement phase transition, the relic density (\ref{eq:lambda0}) appears to effectively ``forget'' the initial conditions in the dark sector~\cite{Carenza:2022pjd,Carenza:2023eua}. Additionally, at temperatures below $T_c$ electromagnetic interactions between the dark glueballs and SM particles can be neglected as their rates fall below the Hubble rate, $\Gamma<H$. Consequently, the dark sector energy in a comoving volume element remains conserved at $T<T_c$ as will be justified later in the next Section. Hence, Eq.~\eqref{eq:lambda0} can also be applied in the other case of two-component DM composed of $a$ and $\eta$ glueballs. Furthermore, the confinement scale in this case remains approximately bounded by Eq.~\eqref{Lambda-const}. We leave a precision analysis of the relic DM abundance in this scenario for a later study. In what follows we, for simplicity, consider a realistic case of comparable contributions of each $a$ and $\eta$ glueball species to the DM abundance in modern Universe as they are produced by a similar dynamics and lead to nearly indistinguishable phenomenological effects. 

\section{Glueball interactions}
\label{sec:GALP}

\subsection{Introductory remarks}

Until now, we have considered a dark $SU(N)$ sector and the dynamics of its simplest composite states: the scalar and pseudoscalar glueballs. This discussion was carried out under the hypothesis of no, or sufficiently weak, interactions with the visible sector. It is instructive to elaborate on glueball phenomenology in more detail despite of weakness of these interactions.

Introducing an interaction between dark gluons and SM particles is not an easy task. Indeed, quarks cannot be coupled directly to an extra unbroken $SU(N)$ gauge group in a renormalizable way. Such an interaction would lead to confinement of a new color charge. Even though a new energy scale associated with this confinement can be smaller compared to the QCD one, having a quark with an extra color charge would show up in measurements of e.g.~the Drell ratio $R_{\mu}$ from electron-positron annihilations~\cite{ParticleDataGroup:2022pth}. As a matter of fact, the electron-positron annihilation process provides a clean environment originally used to measure the number of quark colors in QCD~\cite{Tarnopolsky:1974xb,Litke:1973np}. The ratio between the total inclusive cross section for $e^{+}e^{-}\to \,{\rm hadrons}$ and $e^{+}e^{-}\to\mu^{+}\mu^{-}$ depends on the number of colors and the quark's electromagnetic charges. Any additional quark degree of freedom would enter this parameter, ultimately, in contrast with observations. Thus, we do not invoke a direct renormalizable interaction between visible quarks and dark gluons.

When considering leptons, we exclude any possibility of them being coupled to dark gluons, just because leptons are not confined. In a more intricate model, it would be possible to introduce dark gluon-fermion interactions via a Higgs portal connecting dark fermions to the visible ones. We neglect this possibility for now as that requires a more involved modeling.

While leaving an analysis of the UV complete scenarios featuring different types of dark-to-visible mediation, in this work we adopt a simple and viable EFT approach invoking an interaction between dark gluons and SM gauge bosons via higher-dimensional operators. In order to generate such operators dynamically, one can postulate the existence of e.g.~a heavy fermion, $\Psi$ with mass $M_\Psi$, a single one in the minimal ``dark QCD'' scenario that is charged under both SM and the new $SU(N)$ gauge group. We will discuss possible implications of this scenario below.

\subsection{Glueball-photon interactions}

In the above considered dark QCD framework in the UV regime, a heavy fermion, denoted by $\Psi$ mediates interactions between the visible and dark QCD-like sectors. At lower energies, this framework transitions to an IR effective theory, where an effective interaction between dark gluons and photons emerges through higher-dimensional operators. These operators are radiatively generated at the one-loop level when the fermion $\Psi$ field is integrated out. Consequently, a suppression of the electromagnetic interactions of \(a\) and \(\eta\) is ensured by a heavy fermion scale, \( M_\Psi \gg \Lambda \), as well as by the dark $SU(N)$ coupling found at that energy scale, \( g(\mu=M_\Psi) \). Thus, a large separation in scales effectively decouples the dark glueball dynamics from that of the visible sector in the IR regime of the theory.

In order to introduce highly suppressed electromagnetic interactions into the dark QCD sector, one may consider $\Psi$ carrying an electric charge $q_{\Psi}$ such that the corresponding dark QCD coupling in the UV regime, $g(\mu=M_\Psi) \equiv \epsilon e q_{\Psi}$, appears significantly weaker than its electromagnetic counterpart $e q_{\Psi}$, i.e.~$\epsilon \ll 1$. This condition aligns with the concept of asymptotic freedom in the dark sector. As a result, dimension-8 operators describing the effective dark gluon-photon interactions at energies below $M_\Psi$ take the following form~\cite{Faraggi:2000pv}: 
\begin{equation}
    \mathcal{L} \supset \frac{\epsilon^2\alpha^2}{M_\Psi^4}\,
    \Big[c_\gamma\,G_{\mu\nu}^{a}G^{\mu\nu a} F_{\mu\nu}F^{\mu\nu} + \tilde{c}_\gamma\, G_{\mu\nu}^{a}\tilde{G}^{\mu\nu a} F_{\mu\nu}\tilde{F}^{\mu\nu} \Big]\,,
    \label{eq:naive}
\end{equation}
in terms of the fine structure constant $\alpha$ determined at the matching scale $\mu = M_\Psi$, the electromagnetic field strength tensor, $F^{\mu\nu}$, an its dual, $\tilde{F}^{\mu\nu}$. The dimensionless constants, $c_\gamma$ and $\tilde{c}_\gamma$, can be in principle computed perturbatively. Note, such an EFT description of feeble electromagnetic interactions of dark gluons is viable only in the case of strong scale hierarchy, $\Lambda \ll M_\Psi$.

Applying Eq.~\eqref{GG-to-glue} and retaining only leading terms in $a$ and $\eta$ in the series expansion of the glueball potential around the glueball vacuum, we obtain
\begin{eqnarray*}
    G_{\mu\nu}^{a}\tilde{G}^{\mu\nu a} = \eta_0^3a + 3a_0\eta_0^2\eta \,, \qquad 
    G_{\mu\nu}^{a}G^{\mu\nu a} = -\frac{8g\eta_0^3\,\eta}{\beta(g)}\,.
\end{eqnarray*}
Compared to the above perturbative $g(M_\Psi)$, here $g/\beta(g)$ emerged from the glueball field must be evaluated within the non-perturbative regime of the theory with $g \gg 1$ at energies close to the confinement scale $\mu \sim \Lambda$. Lattice simulations of~\cite{Hasenfratz:2023bok} suggest that at these energies $g/\beta(g) \sim 1$, such that 
the interactions between glueballs and photons are captured to the leading order by the effective Lagrangian
\begin{equation}
    \mathcal{L}_{\rm int} = -\frac{g_{\eta\gamma}}{4} \eta\,F_{\mu\nu}F^{\mu\nu} -\frac{\tilde{g}_{\eta\gamma}}{4} \eta\,F_{\mu\nu}\tilde{F}^{\mu\nu} -\frac{\tilde{g}_{a\gamma}}{4} a \,F_{\mu\nu}\tilde{F}^{\mu\nu}\,.
    \label{eq:lagh}
\end{equation}
One may naturally assume the effective glueball-photon couplings to have comparable strengths suggesting the following compact form \cite{Carenza:2024qaq},
\begin{eqnarray}
     g_{\eta\gamma} &\sim& \tilde{g}_{a\gamma} \sim \tilde{g}_{\eta\gamma} \equiv g_{{\rm GALP}\gamma} \,, \nonumber \\
     g_{{\rm GALP}\gamma} &=& \epsilon^{2}\kappa\,\alpha^{2} \Lambda^{-1}\left( \frac{\Lambda}{M_\Psi}\right)^{4} = \label{eq:gag} \\
     &&\hspace{-0.5cm}=5.3\times10^{-26}\GeV^{-1}\epsilon^{2}\kappa\left(\frac{\Lambda}{100~\eV}\right)^{3} \left(\frac{M_\Psi}{\GeV}\right)^{-4} \,, \nonumber
\end{eqnarray}
in terms of the phenomenologically relevant dimensionless parameter,
\begin{equation}
    \kappa=
    \begin{cases}
  -c_{\gamma}   \frac{8g}{\beta(g)}\left(\frac{\eta_{0}}{\Lambda}\right)^{3}\quad&{\rm for}\quad g_{\eta\gamma}\,,\\
  3\tilde{c}_{\gamma}\frac{a_{0}\eta_{0}^{2}}{\Lambda^{3}}\quad&{\rm for}\quad \tilde{g}_{\eta\gamma}\,,\\
  \tilde{c}_{\gamma}\left(\frac{\eta_{0}}{\Lambda}\right)^{3}\quad&{\rm for}\quad \tilde{g}_{a\gamma}\,,
    \end{cases}
\end{equation}
that can be constrained from the underlined glueball EFT. From astrophysics point of view, it is instructive to consider DM observables that are not sensitive to the CP-nature of the underlined electromagnetic interactions of dark glueballs that is the focus of our current work. Consequently, the glueball states $a$ and $\eta$ behave as axion DM with respect to their electromagnetic interactions outlined in Eq.~\eqref{eq:lagh}. This fundamental property of dark glueballs enables us to call them Glueball Axion-Like Particles (GALPs) in what follows. It is crucial for phenomenology of GALP DM to establish bounds on electromagnetic interactions of GALPs represented by the effective $g_{{\rm GALP}\gamma}$ coupling as introduced above in Eq.~\eqref{eq:gag}.

Due to the apparent similarity between axions and composite GALPs, one may straightforwardly introduce an effective analogue of the Peccei-Quinn (PQ) scale for GALPs as
\begin{equation}
    f_{a}=2.2\times10^{22}~{\rm GeV}\,\epsilon^{-2}\kappa^{-1}\left(\frac{\Lambda}{100~\eV}\right)^{-3}\left(\frac{M_\Psi}{\GeV}\right)^{4}\,.
\end{equation}
Evidently, this scale can extend deeply into the super-Planckian range, reaching magnitudes so large that they are considered unachievable in other traditional axion-like DM models. For instance, in string theory, it is typically constrained to be $f_a \lesssim 10~M_{\rm P}$~\cite{Banks:2003sx,Bachlechner:2014gfa}.
\begin{figure*}[t!]
    \vspace{0.cm}
    \includegraphics[width=0.9\linewidth]{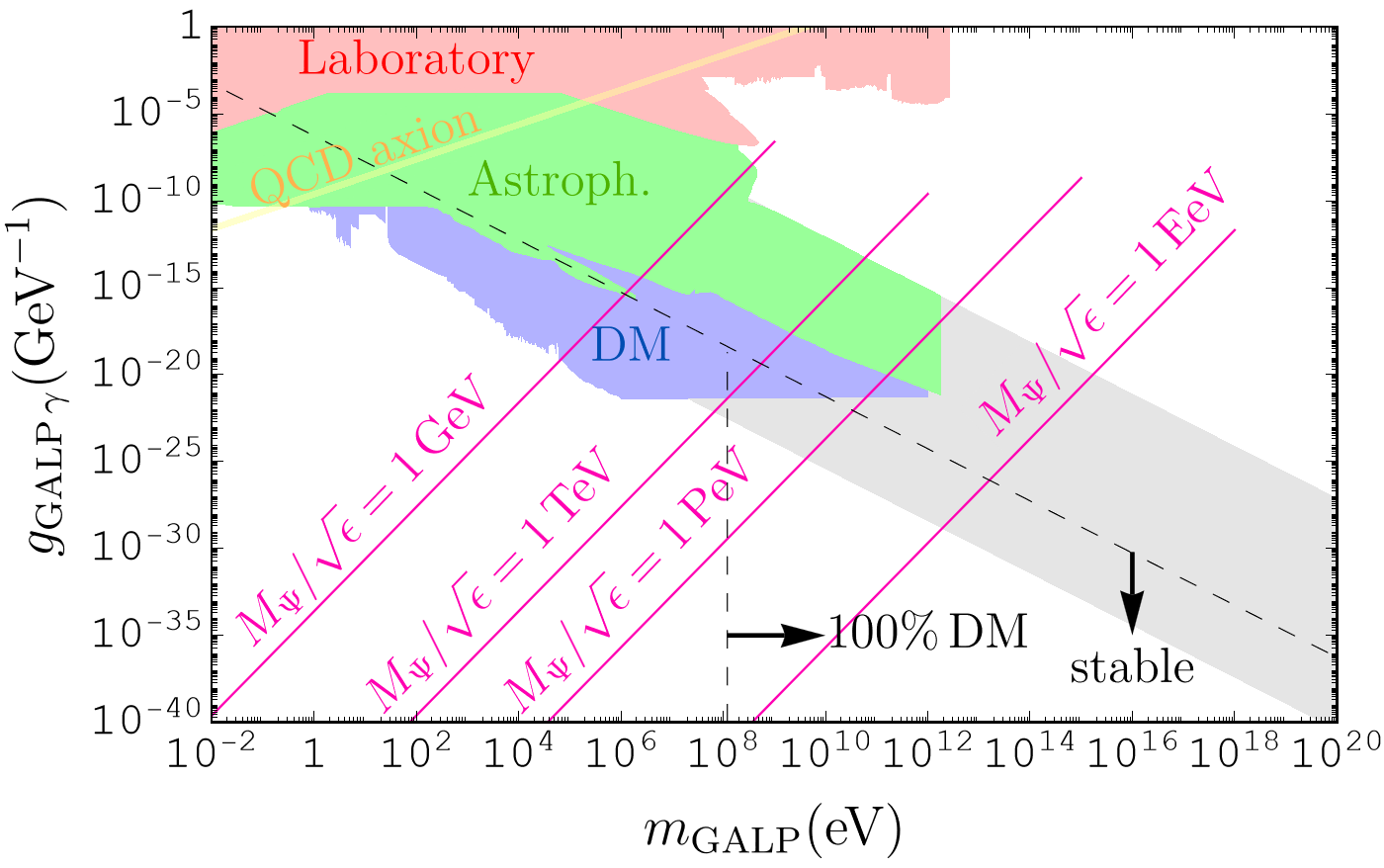}
    \caption{The electromagnetic coupling-mass relation for GALPs considered to be pseudoscalar glueballs $a$, with $m_{\rm GALP}=6\Lambda$ for $N=3$. The colored regions represents laboratory bounds (red), astrophysical constraints that do not assume GALPs to be DM (green) and a combination of  astrophysical and cosmological constraints with the assumption that GALPs provide the totality of DM (blue). Moreover, the gray region shows our estimates of the bounds in the high mass range. For comparison, the canonical QCD axion band is shown in yellow. The violet lines highlight the parameter space consistent with GALP models, for various values of the heavy fermion mass $M_{\Psi}$ and charge $\epsilon$, in terms of $M_{\Psi}/\sqrt{\epsilon}$. The allowed parameter space does not extend indefinitely to high masses because it is limited by the requirement $m_{\rm GALP}<M_{\Psi}$, otherwise the effective description of the GALP-photon interaction in Eq.~\eqref{eq:naive} breaks down. We also delimit the parameter space where GALPs are stable on cosmological timescales; above this line GALPs may exist but do not constitute DM. Finally, if GALPs are stable and constitute the totality of DM, there is a lower limit on their mass due to the self-interacting DM bound, $m_{\rm GALP}\gtrsim 120~\MeV$.}
    \label{fig:paramspace}
\end{figure*}

\subsubsection{Bounds and viable GALP models}

Given the similarity between GALPs and ALPs, it is convenient to summarize their properties in the mass-photon coupling plane, $m_{\rm GALP}$ vs $g_{\rm GALP\,\gamma}$. Since GALP models inherently involve two different particles, with different masses and couplings, the parameter space will refer to just one of the two GALPs -- the lightest one. Figure~\ref{fig:paramspace} shows the GALP parameter space and its constraints, completely analogous to the ALP ones. In this case, $m_{\rm GALP}=6\Lambda$, lighter than twice the other glueball mass and thus composing a fraction of DM, when the decay into photons is not efficient.

The parameter space of GALPs is explored by several arguments. First of all, laboratory bounds (in red) consist in the Light Shining Through the Wall (LSTW) experiments~\cite{DellaValle:2015xxa}, collider searches~\cite{Dolan:2017osp,BESIII:2022rzz,ATLAS:2020hii,Knapen:2016moh} and beam dumps~\cite{CHARM:1985anb,Riordan:1987aw,Blumlein:1990ay,NA64:2020qwq}. A LSTW experiment can probe very light, sub-eV, GALPs by converting part of the photons in a laser beam, into GALPs, via photon-GALP oscillations in an external magnetic field. Due to their weak interactions with the visible sector, GALPs travel unhindered across an optical barrier, the `wall'. Behind this barrier GALPs can reconvert into photons in a magnetic field. In absence of GALPs, a photon signal would not be detected behind the wall. Therefore, a positive signal would point to the existence of GALPs mixed with photons. LSTW have always reported exclusion limits on the GALP properties, given that no positive signal was detected. More massive GALPs, in the MeV-GeV range, can be produced in laboratory searches through energetic particle interactions. Once those GALPs are produced, beam dump experiments look for signatures of their decay. This approach allows one to probe very massive GALPs with short lifetimes.

Indirect searches collect a large number of astrophysical and cosmological observations, to unveil GALP properties. For example, astrophysical searches assuming GALPs to constitute the totality of DM are in blue (these constraints should be rescaled by a factor $\sqrt{f}$, where $f$ is the DM fraction in GALPs, however this scaling would amount to a small change)~\cite{Cadamuro:2011fd,Grin:2006aw,Todarello:2023hdk,Yin:2024lla,Janish:2023kvi,Carenza:2023qxh,Nakayama:2022jza}. Unstable GALP DM would decay into observable photons across a vast photon energy range. By using several astrophysical photon data, it is possible to search for lines caused by GALP decay into two photons. Similar searches can be performed on extra-galactic scales, by considering the decay of all GALPs on cosmological scales, producing a non-monochromatic spectrum.

In green, we show the searches that do not rely on the assumption of GALPs to be DM~\cite{Ayala:2014pea,Caputo:2022mah,Diamond:2023cto,Depta:2020wmr,Langhoff:2022bij}. In this class of bounds one finds stellar limits, given by the production of GALPs at different stellar stages. The energy lost into GALPs has observable effects in altering the stellar evolution. Moreover, GALPs produced in stars might decay into photons leading to observable signals that can be compared to historical data, such as the gamma-ray measurements of Solar Maximum Mission at the time of SN 1987A. In this set of searches, we also include the irreducible contribution of GALPs from freeze-in, establishing a population in the early Universe, and then decaying into photons today.

All the colored bounds refer to published data, and we extend these exclusion regions to higher masses with very simple considerations (gray area). First of all, the upper limit of the gray region is determined by GALPs decaying into photons during the Big-Bang Nucleosynthesis (BBN) and affecting the primordial nuclei abundances. Note that GALPs that are too short-lived will not affect the observables because their decay would happen much before BBN, therefore the bounds do not extend indefinitely to large coupling. This bound is prolonged from the one in Ref.~\cite{Depta:2020wmr}, and it should still be valid also at higher masses. This is true, as long as the GALP mass is below the reheating temperature, as assumed in this case. The lower limit of the gray region is related to DM searches for ultramassive particles. In Refs.~\cite{Blanco:2018esa,Munbodh:2024ast} it is shown that very massive DM, up to $10^{20}$~eV, decaying into various possible channels is excluded if the decay rate is $\sim\mathcal{O}(10^{-26}{\rm s}^{-1})$, at least. Also for DM decaying into photons, at such high energies, the prompt signal will induce a cascade of various less energetic decay products, similarly to the mentioned analyses. Therefore, we set a rough limit by requiring the GALP to have a decay rate equal to $10^{-26}{\rm s}^{-1}$. The gray area shown here is just a very rough estimate of the bounds, whose accurate calculation is left for the future. 

In Fig.~\ref{fig:paramspace} we denote the portion of the parameter space allowing for GALPs stable on cosmological timescales against the decay into photons, the arrow labeled `stable'. Moreover, if GALPs constitute the DM, there is a lower limit on their mass due to self-interactions $m_{\rm GALP}\gtrsim180~\MeV$~\cite{Carenza:2023eua}, indicated by the arrow labeled `$100\%$~DM'. The important message of this plot is the parameter space available for consistent GALP models. The violet lines denote the value of $g_{a\gamma}$ obtained in a GALP model where the photon interaction stems from a fundamental dark gluon-photon interaction, as in Eq.~\eqref{eq:naive}, suppressed by a UV cutoff scale. As was discussed above, the latter can be associated with the mass of a mediator $M_\Psi$ between dark gluons and photons, and its value is largely unconstrained. In this plot, we show the result of Eq.~\eqref{eq:gag}. For the consistency of the interaction in Eq.~\eqref{eq:naive} we require that $m_{\rm GALP}\lesssim M_{\Psi}$, otherwise this effective interaction is not applicable anymore. As $M_{\Psi}$ grows, the viable GALP becomes more massive and weakly coupled. We note that in this model a stable GALP composing the totality of DM can be obtained in models with $M_{\Psi}\gtrsim 1~\TeV$, and a coupling low enough to evade decay constraints. 

\subsubsection{Summary of photophilic GALPs}

As discussed above, GALPs coupled to photons are an interesting paradigm for ALP model building. In this model, it is natural to predict a very massive GALP, with masses well above the GeV scale. The coupling to photons, in sharp contrast with canonical ALP models, can be extremely suppressed by introducing GALP-photon interactions mediated by a heavy degree of freedom with mass above the TeV scale. The defining properties of this model make possible to achieve a suppression of the GALP-photon coupling analogous to a super-Planckian PQ scale. Indeed, super-Planckian values are obtained, for example, when $M_{\Psi}=1\,{\rm PeV}$ and $\Lambda\ll 100$~GeV. Note that these energy scales have an order of magnitude of typical energy scales in particle physics. 

Thus, this model makes it possible to introduce extremely (Planck)-suppressed interactions, by postulating new physics at the TeV-PeV scale. The phenomenology of GALPs in this range is not explored in detail given that ALPs with these properties were considered not achievable. However, we stress that this work proves that such models are viable and relatively easy to realize in a composite scenario. For future studies, this portion of the parameter space should be explored with all its possible phenomenological consequences.
\begin{figure*}[t!]
    \vspace{0.cm}
    \includegraphics[width=0.9\linewidth]{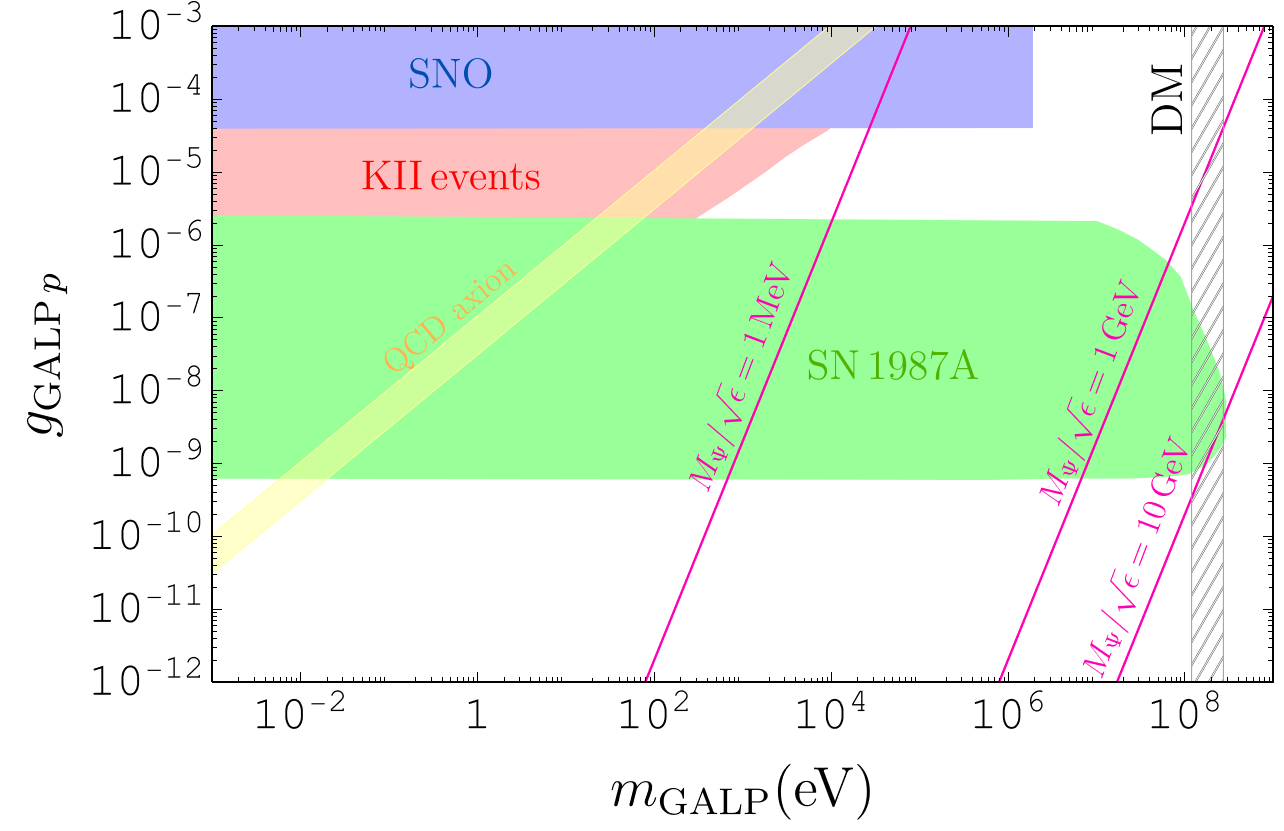}
\caption{Parameter space of GALPs (corresponding to a pseudoscalar glueball) expressed in terms of its mass $m_{\rm GALP}$, related to the confinement scale $\Lambda$ by $m_{\rm GALP}=6\Lambda$, and the GALP-proton coupling $g_{{\rm GALP}\,p}$. Here, we assume no coupling to neutrons. The colored regions represents the bound from SN 1987A, due to cooling (green) and absence of events in KII (red). The SNO bound is shown in blue. The vertical hatched area labeled as `DM', shows the narrow mass range in which GALPs are completely stable against decay into two pions, and they can constitute the totality of DM. The canonical QCD axion band is shown in yellow. The violet lines show the parameter space consistent with GALP models, for various values of the heavy fermion mass $M_{\Psi}$ and charge $\epsilon$, in terms of $M_{\Psi}/\sqrt{\epsilon}$. The allowed parameter space does not extend indefinitely to high masses because it is limited by the requirement $m_{\rm GALP}<M_{\Psi}$, otherwise the effective description of the GALP-gluon interaction in Eq.~\eqref{eq:naivegluon} breaks down. }
\label{fig:paramspacegap}
\end{figure*}

\subsection{Glueball-gluon interactions}

Also, the dark gluon-QCD gluon interaction can be introduced similarly to the one with photons discussed in the previous Section, i.e.
\begin{eqnarray} \nonumber
    &&\mathcal{L} \supset \frac{\epsilon^2\alpha_{s}^2}{32\pi^{2}M_\Psi^4}\,
    \Big[c_g\,G_{\mu\nu}^{a}G^{\mu\nu a} F^{a}_{\mu\nu}F^{a\mu\nu} \\ && \qquad +\; \tilde{c}_g\, G_{\mu\nu}^{a}\tilde{G}^{\mu\nu a} F^{a}_{\mu\nu}\tilde{F}^{a\mu\nu} \Big]\,,
    \label{eq:naivegluon}    
\end{eqnarray}
where $\alpha_{s}$ represents the QCD strong coupling, $F^{a}_{\mu\nu}$ is the gluon field strength tensor and $\tilde{F}^{a\mu\nu}$ is its dual. We will work under the assumption that the dark sector confines before the QCD sector does in the early Universe, which happens at the temperature $\Lambda_{\rm QCD}\simeq 200~\MeV$. This condition can be written as $\Lambda\gtrsim\zeta_{T}^{-1}\Lambda_{\rm QCD}$, allowing for a very cold dark sector, i.e.~$\zeta_{T}^{-1}\ll1$, with a relatively low confinement scale. Note that the considered interaction would, in general, not allow the glueballs to address the strong-CP problem because it is not shift-symmetric in the initial Lagrangian in Eq.~\eqref{eq:lagrangian}. Thus, glueballs may not be treated as composite axions, but they act as ALPs.

After the confinement of the dark sector, the induced interactions, at the lowest order, read
\begin{equation}
    \mathcal{L}_{\rm int} = \frac{\alpha_{s}^{2}}{32\pi^{2}}\Bigg[\frac{\eta}{f_{\eta}}\,F^{a}_{\mu\nu}F^{a\mu\nu} +\frac{\eta}{\tilde{f}_{\eta}}\,F^{a}_{\mu\nu}\tilde{F}^{a\mu\nu} +\frac{a}{\tilde{f}_{a}} \,F^{a}_{\mu\nu}\tilde{F}^{a\mu\nu}\Bigg]\,,
    \label{eq:laghgluon}
\end{equation}
where we introduced the effective PQ scales
\begin{eqnarray}
     f_{\eta} &\sim& \tilde{f}_{a} \sim \tilde{f}_{\eta} \equiv f_{{\rm GALP}} \,, \nonumber \\
     f_{{\rm GALP}} &=&\Lambda \,\epsilon^{-2}\kappa ^{-1}\left( \frac{M_\Psi}{\Lambda}\right)^{4} = \label{eq:gap} \\
     &&\hspace{-0.5cm}=10^{21}\GeV\,\epsilon^{-2}\kappa^{-1}\left(\frac{\Lambda}{100~\eV}\right)^{-3} \left(\frac{M_\Psi}{\GeV}\right)^{4} \,, \nonumber
\end{eqnarray}
Therefore, the low-energy manifestation of the GALP-gluon coupling is an effective interaction with nucleons, in analogy to the QCD axion models. The GALP-proton coupling for the field $a$ can be calculated in the chiral perturbation theory leading to the following expression~\cite{GrillidiCortona:2015jxo}
\begin{equation}
    \begin{split}
        g_{{\rm GALP}p}&=-0.47\frac{m_{N}}{f_{{\rm GALP}}}=\\
        &=-4.4\times10^{-22}\epsilon^{2}\kappa\left(\frac{\Lambda}{100~\eV}\right)^{3} \left(\frac{M_\Psi}{\GeV}\right)^{-4}\,.
    \end{split}
    \label{eq:gapc}
\end{equation}
It is more difficult to give an estimate for the scalar component $\eta$, featuring both scalar and pseudoscalar interactions with gluons, even though the order of magnitude of the coupling would be comparable to that in Eq.~\eqref{eq:gapc}. Despite the smallness of this coupling, it has a steep dependence on the confinement scale, which can be much higher than $100$~eV for models with a very cold dark sector required to confine at scales higher than the QCD one.

\subsubsection{Bounds and viable GALP models}

In Fig.~\ref{fig:paramspacegap} we summarize the constraints in the $g_{{\rm GALP}\,p}$-$m_{\rm GALP}$ plane, and the parameter space allowed by GALP models. GALPs coupled to nuclear matter can be copiously produced inside Supernovae (SNe) and escape the star due to their weak interactions stealing energy from the core. When the energy lost in this form is comparable with the standard neutrino losses, the SN neutrino burst will be shortened, ultimately being in contrast with observations of SN 1987A. Following this argument, the cooling bound from SN 1987A is marked in green~\cite{Lella:2023bfb}. 

The same GALP flux produced in SN 1987A explosion might have been detectable in terrestrial neutrino detectors, such as Kamiokande-II (KII)~\cite{Lella:2023bfb,Carenza:2023wsm}, via GALP absorption on oxygen nuclei and their subsequent radiative de-excitation. Despite the sparseness of SN 1987A neutrino data, no anomalies were found pointing to such an exotic signal. Therefore, in red we show the bound on the absence of GALP-induced events in KII. 

In addition to SNe, GALPs coupled to nucleons can be produced in the solar interior, and then detected on the Earth. The blue region is excluded by the search for $p+d\to \, ^3 {\rm He}+a\,(5.5\,{\rm MeV})$ solar GALP flux in the Sudbury Neutrino Observatory~\cite{Bhusal:2020bvx}. Similarly to the GALP model, the yellow band refers to the standard QCD axion models~\cite{GrillidiCortona:2015jxo}. The violet lines denote viable projections of the parameter space in GALP models with different high-energy cutoff scales values $M_\Psi$.

\subsubsection{Summary of nucleophilic GALPs}

Nucleophilic GALPs are an interesting possibility to constitute a heavy ALP model that can be probed in astrophysical searches. Currently, there is a limited interest in heavy ALPs coupled to nucleons, and we hope this model can strengthen the case for such studies.

Note that, in principle, GALPs coupled to QCD might decay into two pions (as parity is not conserved) making them unstable on cosmological timescales, as soon as the decay process is kinematically allowed. A more precise calculation of this decay rate would be needed, but we expect that for extremely small $g_{{\rm GALP} \,p}$ GALP DM is stable on cosmological scales, and its hadronic decay might lead to very peculiar indirect signatures.

Nevertheless, there is a narrow mass range $180~\MeV\lesssim m_{a}\lesssim 270~\MeV$, denoted by the hatched band in Fig.~\ref{fig:paramspacegap}, in which GALPs are completely stable against decays and can constitute the totality of DM. Interestingly, a portion of this parameter space is probed by the high-mass end of the SN 1987A cooling bound.

\section{GALP mass-coupling relation}
\label{sec:relation}

As discussed earlier, the GALP model most relevant to DM phenomenology is the scenario in which GALPs interact with photons. Below, we establish a relationship between the GALP mass and its coupling to photons, assuming that GALPs account for the entirety of DM. In particular, the physical parameters of the GALP EFT outlined above such as the electromagnetic coupling, $g_{{\rm GALP}\gamma}$, the mass scale, $m_{\rm GALP}$, as well as the GALP relic abundance, can be connected to each other. In the UV limit of the underlined dark QCD, $T\gg M_{\Psi}$, the relativistic $\Psi$ fermions are in thermal equilibrium with the photon gas with comparable densities, $\rho_{\gamma}\sim\rho_{\Psi}$. At lower temperatures, $T\sim M_{\Psi}$, the $\Psi$-annihilation reactions into dark gluon and photon final states such as $\Psi\bar{\Psi}\to \tilde{g}\tilde{g}$, $\Psi\bar{\Psi}\to \tilde{g}\gamma$, $\Psi\bar{\Psi}\to \gamma\gamma$ become efficient. At those energy scales, the annihilation processes into photons dominate since $\epsilon\ll 1$, hence increasing the $\gamma$-density relative to that of dark gluons. Indeed, the latter gets suppressed by $\epsilon^2$-factor compared to the visible radiation density leading to a significant temperature difference between the dark and visible sectors.
In particular, at temperatures right below the $\Psi$ mass scale, $T\equiv T_\gamma \simeq M_\Psi$, one finds
\begin{equation}
\rho_{\gamma}\sim 2\rho_{\Psi}\propto T_\gamma^4\,,\qquad
\rho_{\tilde{g}}\sim\frac{1}{2}\epsilon^2\rho_{\Psi}\propto (N^2-1)T_{\tilde{g}}^4\,, \label{densities}
\end{equation}
in terms of photon $T_\gamma$ and dark gluon $T_{\tilde{g}}$ temperatures, and with $N^2-1$ dark gluon colors. Besides, we took into consideration that $\tilde{g}$ density is mostly impacted by the $\Psi\bar{\Psi}\to \tilde{g}\gamma$ reaction such that only half of the initial $\Psi$ radiation density gets transferred into the $\tilde{g}$ density in the cosmological plasma. The role of $\Psi\bar{\Psi}\to \tilde{g}\tilde{g}$ process is sub-leading due to an extra $\epsilon^2$ suppression factor and can therefore be discarded in this approximate consideration. 
\begin{figure*}[t!]
    \vspace{0.cm}
    \includegraphics[width=0.99\linewidth]{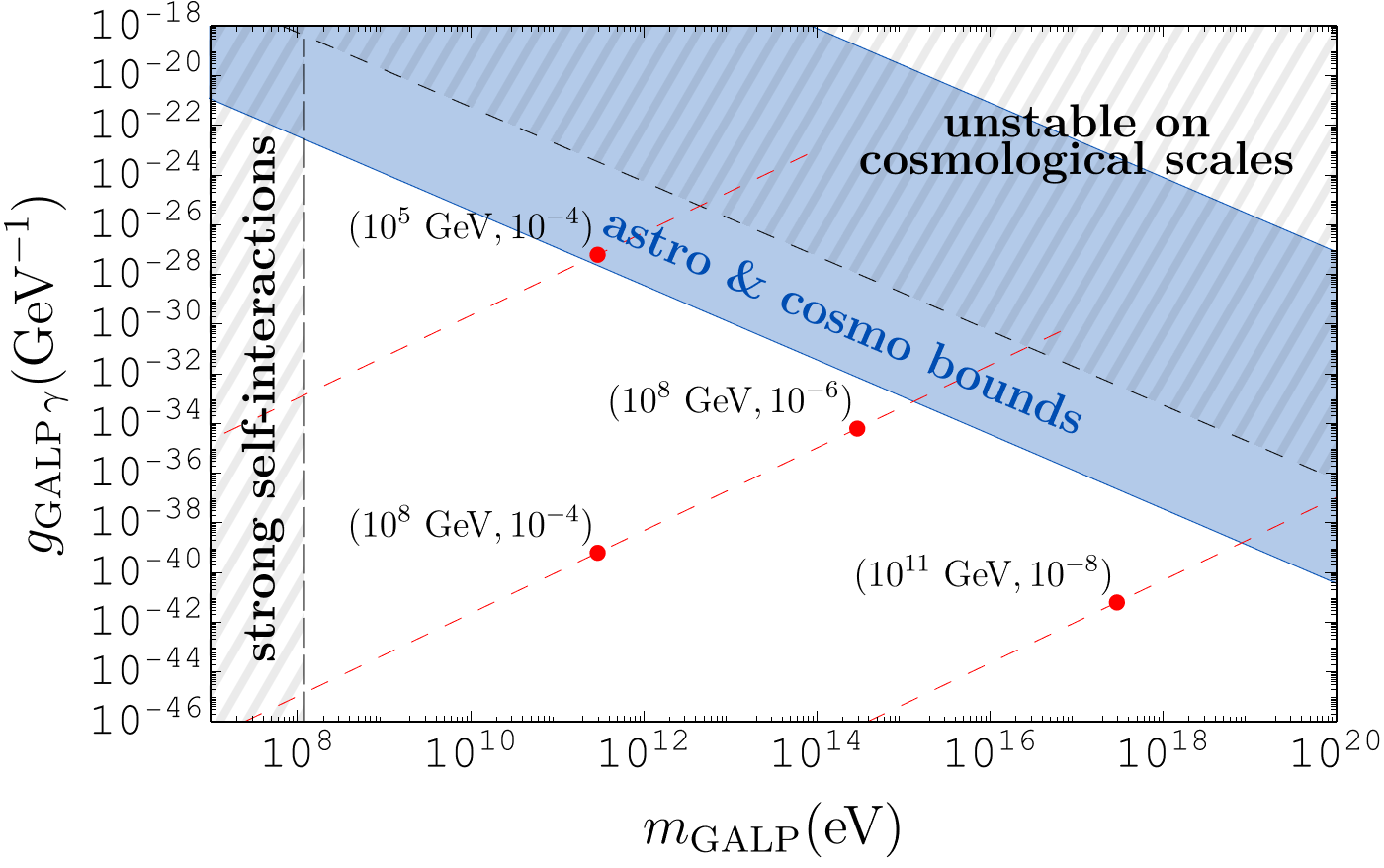}
    \caption{The electromagnetic coupling-mass relation for GALP DM, considering $m_{\rm GALP}=6\Lambda$ for $N=3$, as well as $\Lambda_{0}=133$~eV~\cite{Carenza:2023eua}. Cosmological and astrophysical measurements rule out the GALP parameter space in the blue region, while domains with no viable GALP DM candidates are highlighted by hatched areas, either due to instability above the diagonal dashed line or by means of the DM self-interaction constraints that are violated for light GALPs. Different GALP scenarios parameterised by $(M_\Psi,\epsilon)$ pairs are indicated by red dots. The GALP-$\gamma$ coupling $g_{{\rm GALP}\gamma }$ specified in Eq.~\eqref{eq:relationnew} is given as a function of $\epsilon$ by red dashed lines for distinct values of the heavy fermion mass scale $M_\Psi$. Figure taken from Ref.~\cite{Carenza:2024qaq}.}
    \label{fig:models}
\end{figure*}

Considering that entropy production in the visible sector impacts only $T_\gamma$, we deduce for the dark gluon-to-photon temperature ratio 
\begin{equation}
    \zeta_{T}^{-1}\simeq \left(\frac{g_{*,s}(T^{(0)}_{\gamma})}{g_{*,s}(M_\Psi)}\right)^{1/3}\left(\frac{\epsilon^{2}}{4(N^{2}-1)}\right)^{1/4} \propto \sqrt{\epsilon} \ll 1 \,,
    \label{eq:2}
\end{equation}
in terms of the number of entropic $\gamma$ d.o.f.'s $g_{,s}(T^{(0)}_{\gamma}) = 3.909$ in the modern Universe at $T = T^{(0)}_{\gamma}$, whereas the corresponding quantity at the $\Psi$ annihilation epoch is taken to be $g_{,s}(M_\Psi) = 106.75$. We remind that this parameter is critical for GALP-dominated DM in the Universe today ensured by $\zeta_{T}^{-3} \Lambda / \Lambda_0 \simeq 1$ according to Eq.~\eqref{eq:lambda0}. The latter therefore effectively establishes both the dark confinement scale $\Lambda$ and the universal GALP mass scale taken to be equal $m_{\rm GALP} \simeq 6\Lambda$ for $SU(3)$ theory (in general, the relation between confinement scale and glueball mass can be calculated via lattice simulations; see e.g.~\cite{Curtin:2022tou}). Assuming that all DM consists of GALPs, it is straightforward to establish the GALP mass-coupling relation for $\epsilon \ll 1$ and $N = 3$ in conjunction with Eq.~\eqref{eq:gag} and \eqref{eq:2},
\begin{equation}
    g_{{\rm GALP}\gamma } \simeq \frac{ 10^{-(21.1\pm0.4)}}{\GeV}\kappa\left(\frac{m_{\rm GALP}}{\keV}\right)^{\frac{5}{3}}\left(\frac{M_\Psi}{\GeV}\right)^{-4}\,,
    \label{eq:relationnew}
\end{equation}
where the uncertainties stem from those of the relic DM density analysis \cite{Carenza:2024qaq}. The above expression is analogical to the coupling-mass relation for ordinary QCD axions, c.f.~\mbox{$g_{a\gamma} = 2 \times 10^{-10} \GeV^{-1} (m_{a} / \eV)$}~\cite{GrillidiCortona:2015jxo}. Indeed, it contains $M_\Psi$ scale as a free parameter, while in this work we take $\kappa \sim 1$, for simplicity. The GALP-$\gamma$ coupling gets more suppressed with an increase of $M_\Psi$ scale yielding a phenomenologically consistent family of GALP scenarios for DM. 

In Fig.~\ref{fig:models} we present the parameter space for GALP DM. GALPs with sufficiently high masses $m_{\rm GALP}\gtrsim 120$~MeV, avoid self-interaction constraints\cite{Smirnov:2019ngs} and they remain stable on cosmological timescales for low enough values of $g_{{\rm GALP}\gamma}$ (shaded regions are excluded). The blue region provides an upper bound on $g_{{\rm GALP}\gamma}$, whose upper limit is found from cosmological constraints while astrophysical DM searches place its lower limit. More specifically, for larger $g_{{\rm GALP}\gamma}$, the GALP DM decays into $\gamma$ in the course of BBN epoch may spoil abundances of primordial nuclei~\cite{Depta:2020wmr}. For the considered parameter space regions, the bound of Ref.~\cite{Depta:2020wmr} has been extended for larger $m_{\rm GALP}$ while keeping it below the reheating temperature. On the other hand, as was shown by Refs.~\cite{Blanco:2018esa,Munbodh:2024ast}, the searches for very heavy DM particles with mass up to $10^{20}$~eV impose bounds on their decay rates down to about $\sim\mathcal{O}(10^{-26}{\rm s}^{-1})$, approximately.

Besides, in the same figure, by means of red dots we indicate a family of GALP DM scenarios parameterized by different values of $M_\Psi$ and $\epsilon$, while red dashed lines show $g_{{\rm GALP}\gamma}$ given by Eq.~\eqref{eq:relationnew} as a function of $\epsilon$ for a fixed $M_\Psi$ found in parenthesis $(M_\Psi,\epsilon)$. Such a function exhibits a drift of GALP models to lower values of $g_{{\rm GALP}\gamma}$ and $m_{\rm GALP}$ as $\epsilon$ increases. While the GALP mass can be as low as the QCD scale, $m_{\rm GALP}\sim 120$~MeV found at $\epsilon\sim 2\times 10^{-2}$, the heaviest GALPs can reach masses as high as $m_{\rm GALP}\simeq M_\Psi$ being the upper limit.

It is worth noticing here that the GALP model involves heavy DM states effectively circumventing the unitarity constraints~\cite{Griest:1989wd}. In fact, the unitarity bound plays a crucial role in DM freeze-out scenarios that assume the DM particles being in thermal equilibrium with the visible sectors at very early times in the cosmological history. However, the considered composite GALP DM has not generated through freeze-out processes. Instead, they emerged due to confinement in a dark gauge sector at scales $\Lambda$ being way below the energy scale $M_\Psi$ at which dark QCD has been in a state of equilibrium with the SM bath. Furthermore, the GALPs' self-interaction cross section governed by $m_{\rm GALP}$ is ensured to be always below the geometrical cross section~$\sigma_{\rm geom}\sim 1/m_{\rm GALP}^{2}$~\cite{Smirnov:2019ngs}.

To summarise, we find vast parameter space domains corresponding to heavy DM GALP models with not-so-small couplings $g_{{\rm GALP}\gamma}$ that have not yet been probed or excluded by existing measurements as indicated by Fig.~\ref{fig:models}, thus, suggesting intriguing phenomenological opportunities. One may anticipate that indirect DM searches~\cite{Cirelli:2010xx,Cirelli:2024ssz} will be sensitive to these regions of the GALP parameter space, particularly to those close to the blue domain. While further more dedicated studies of the viable parameter space as well as the existing and future bounds are necessary, our analysis suggests that yet available GALP parameter space can be obtained in a dark QCD-like $SU(N)$ gauge theory with small gauge coupling $g$ being $4-8$ orders of magnitude weaker than the corresponding electromagnetic coupling and with a single heavy $\Psi$ state in the mass range above a TeV scale. Consequently, this framework represents a minimal model in which GALP DM properties can be comprehensively characterized.

In conclusion, this work illustrates a method for deriving composite axion-like particles featuring the properties distinct from those in existing models. While not resolving the strong-CP problem, it is crucial to highlight that GALPs provide a natural framework to encourage investigations into very massive ALP DM -- a research field that remains largely uncharted in the literature (see Refs.~\cite{Ghosh:2022rta,Ghosh:2023xhs,Ghosh:2024boo}). Furthermore, our novel GALP framework enables to effectively achieve extremely super-Planckian PQ scales. With these unique attributes, we expect a growing interest and research efforts in exploring GALPs DM phenomenology and their detection prospects filling the gap between the axion and WIMP DM studies.

\section{Main differences between ALPs and GALPs}
\label{sec:differences}

Here, we briefly summarize the main innovations and differences introduced by GALPs into the ALP paradigm.

\begin{itemize}
\item \textbf{Not a composite axion.} The proposed GALP model is not leading to a solution of the strong-CP problem, as done by composite axion models~\cite{Kim:1984pt,Choi:1985cb,Redi:2016esr,Gavela:2018paw}. 
\item\textbf{No scale hierarchy problem.} Composite axion models are usually invoked to explain why the expected PQ scale $f_{a}\sim(10^{9}-10^{10})~\GeV$ for DM axions is so much larger than the electroweak scale and, at the same time, significantly lower than the Grand Unification one. In this model, we show that the effective PQ scale $f_{a}$ can have a large variability, even reaching super-Planckian values for $M\gtrsim\TeV$ (see Eq.~\eqref{eq:gap}). 
\item \textbf{DM more massive than usual axion models.} GALP DM has a lower limit on its mass, $m_{\rm GALP}\gtrsim 120~\MeV$, making GALPs a paradigm more similar to traditional Weakly Massive Interacting Particles, than the usual light axion DM. This is an interesting feature of the GALP model, motivating the DM searches in a vast region of the parameter space.
\item \textbf{Avoid unitarity bounds.} Our model avoids the upper bound on the DM mass imposed by unitarity~\cite{Griest:1989wd}. Indeed, the latter requires DM to be thermally produced, i.e.~to have been in thermal equilibrium with the visible sectors in the very early Universe, while the considered GALPs have actually never been in equilibrium. Moreover, the self-interaction GALP cross section satisfies the requirement of being below the geometrical one, $\sigma_{\rm geom.}\sim 1/m_{\rm GALP}^{2}$~\cite{Smirnov:2019ngs}, for vast regions of the parameter space.
\item \textbf{Intriguing phenomenology.} The GALP model requires to study the phenomenology of heavy and, possibly, strongly interacting ALPs that acquired interest in recent years, from a model-independent fashion, e.g.~\cite{Knapen:2016moh,Pilaftsis:2015ycr,Kelly:2020dda,Lucente:2020whw,Carenza:2020zil,Lucente:2022wai,Diamond:2023cto,Diamond:2023scc,Caputo:2022mah,Muller:2023pip,Dev:2023hax,Lella:2023bfb,Muller:2023vjm,Lella:2022uwi,Calore:2021klc,Caputo:2021kcv,Ahmadvand:2023lpp,Giannotti:2010ty,Mori:2021apk,Bauer:2017ris,Balkin:2023gya,Mosala:2023sse}. We find a physical motivation for this phenomenology in the GALP model.
\item \textbf{An extended dark sector.} If the term `GALP' refers to the pseudoscalar glueball, any GALP model which requires the existence of the scalar glueball can be searched for in parallel with the GALPs. A joint signal would be a strong hint towards such a two-component DM model.
\end{itemize}

\section{Conclusions}
\label{sec:conclusions}

In this study, we have investigated the implications of introducing pseudoscalar glueballs into the dark YM sector framework, extending previous analyses of scalar glueballs as DM candidates~\cite{Carenza:2022pjd,Carenza:2023eua}. We have derived an effective Lagrangian determining the dynamics of both scalar, $\eta$, and pseudoscalar, $a$, glueballs. As expected, two glueball states strongly interact with each other allowing for both scattering and decay of the heavier pseudoscalar glueball into two scalar ones, $a\to \eta\eta$. If such a decay mode is allowed kinematically, all the DM would be composed by the lightest glueball $\eta$ only; otherwise, both $a$ and $\eta$ may be present as components of the DM today.

In this work, we considered the possibility of glueballs interacting feebly with SM particles. In particular, starting from a fundamental dimension-8 operator tying dark gluons and SM gauge bosons (photons and gluons), we derived interactions of the composite glueball states with photons and nucleons, with the latter being induced at low energies by the gluon coupling. This exploration revealed an intriguing similarity between these glueball models and ALPs. Therefore, we introduced the concept of GALPs to describe glueballs weakly coupled to the SM. In this preliminary study, we estimated the parameter space viable for GALPs, in particular, motivating phenomenological searches for heavy GALPs.

We briefly discussed a toy model in which the portal between SM and GALPs is responsible for GALP production in the early Universe, relating the DM temperature and abundance, to the GALP-photon coupling. In this model, we strongly reduced the number of free parameters of the theory predicting a mass-coupling relation for GALPs, highlighting them as an alternative DM model to the traditional axion and ALP models.

The study of GALPs opens up new avenues for research, particularly in understanding their interactions with SM particles and their implications for experimental searches. Future investigations could explore the constraints on GALP parameter space and devise experimental strategies to probe their existence. In particular, it would be interesting to probe in detail the models for various types of portal interactions between dark gluons and SM particles, in order to understand the dark gluon thermalization, the associated DM abundance and how the constraints are impacted.

In conclusion, our work highlights the richness of dark YM sectors and their potential role in elucidating the nature of DM. By considering both scalar and pseudoscalar glueballs, as well as their interactions with the SM, we contribute to a more comprehensive understanding of the glueball DM paradigm and pave the way for further exploration in this intriguing field of study.

\acknowledgements
We warmly thank John March-Russell for fruitful discussions. This article/publication is based upon work from COST Action COSMIC WISPers CA21106, supported by COST (European Cooperation in Science and Technology). The work of PC is supported by the European Research Council under Grant No.~742104 and by the Swedish Research Council (VR) under grants  2018-03641, 2019-02337 and 2022-04283.

\appendix

\section{Details of the glueball potential}
\label{app:details}

In this appendix, we discuss more in detail how to connect the five microscopic parameters appearing in the potential of Eq.~\eqref{eq:potential}, $c_{i}$ for $i=0,\dots,4$ plus the two parameters $\eta_{0}$ and $a_{0}$, to observable quantities. This step is necessary in order to draw physical conclusions from such an EFT potential and to fix its parameters in a physically sensible way. Here we will assume that $f(\mathcal{A}/\mathcal{H})=0$. Expanding the potential in Eq.~\eqref{eq:potential} to the second order in each field around the minimum $(\eta_{0},a_{0})$, we obtain:
\begin{equation}
    \begin{split}
  V_{\rm eff}&\simeq \frac{m_{\eta}^{2}}{2} \eta^{2}+\frac{m_{a}^{2}}{2} a^{2}+\lambda_{11} \Lambda^{2}\eta\, a+\\
    &+\lambda_{22} \eta^{2} a ^{2}+\frac{\lambda_{21}}{\Lambda} \eta^{2} a+\frac{\lambda_{12}}{\Lambda}  \eta\, a^{2}+\\
&+V_{0}+\lambda_{10} \Lambda^{3}\eta+\lambda_{01}\Lambda^{3} a+\dots\,,
    \end{split}
    \label{eq:expapp}
\end{equation}
where $\eta=\mathcal{H}-\eta_{0}$ and $ a=\mathcal{A}-a_{0}$, and we neglected higher order terms. The terms in the last line of Eq.~\eqref{eq:expapp} vanish due to the minimization condition. Moreover, we set the value of the potential in the minimum to zero, i.e.~$V(\eta_{0},a_{0})\equiv V_{0}=0$. These three conditions read
\begin{equation}
    \begin{split}
  V_{0}&=  c_{0}\eta_{0}^{4}\ln\left(\frac{\eta_{0}}{\Lambda}\right)+\frac{\theta}{4}a_{0}\eta^{3}_{0}+c_{1}\eta_{0}^{4}+\\
  &\quad+c_{2}\eta_{0}^{2}a_{0}^{2}+c_{3}a_{0}^{4}+c_{4}\eta_{0}a_{0}^{3}=0\,,\\
  \lambda_{10}\Lambda^{3}&=4c_{0}\eta_{0}^{3}\ln\left(\frac{\eta_{0}}{\Lambda}\right)+\frac{3\theta}{4}\eta_{0}^{2}a_{0}+\\
  &\quad+(c_{0}+4c_{1})\eta_{0}^{3}+2c_{2}\eta_{0}a_{0}^{2}+c_{4}a_{0}^{3}=0\,,\\
  \lambda_{01}\Lambda^{3}&=\frac{\theta}{4}\eta_{0}^{3}+2c_{2}\eta_{0}^{2}a_{0}+4c_{3}a_{0}^{3}+3c_{4}\eta_{0}a_{0}^{2}=0\,.\\
    \end{split}
      \label{eq:V0app}
\end{equation}

The glueball masses are defined in terms of the microscopic parameters as
\begin{equation}
    \begin{split}
        \frac{m_{\eta}^{2}}{2}&=\frac{7}{2}c_{0}\eta_{0}^{2}\left[1+\frac{12}{7}\ln\left(\frac{\eta_{0}}{\Lambda}\right)\right]+\\
        &\quad+\frac{3\theta}{4}\eta_{0}a_{0}+6c_{1}\eta_{0}^{2}+c_{2}a_{0}^{2}\,,\\
     \frac{m_{a}^{2}}{2}&=c_{2}\eta_{0}^{2}+6c_{3}a_{0}^{2}+3c_{4}\eta_{0}a_{0}\,,
    \end{split}
    \label{eq:massapp}
\end{equation}
and they are observables of the model. Therefore, fixing them will lead to two more conditions on the microscopic parameters of the theory. At the lowest order, the mass mixing between the two glueball states can be rotated away. Therefore, the observable glueball states will be mass eigenstates. This corresponds to set the linear coupling between the glueball fields to zero, giving a sixth constraint
\begin{equation} 
\lambda_{11}\Lambda^{2}=\frac{3\theta}{4}\eta_{0}^{2}+4c_{2}\eta_{0}a_{0}+3c_{4}a_{0}^{2}=0\,,\\
\label{eq:linapp}
\end{equation}
In addition, the potential in Eq.~\eqref{eq:expapp} at the quadratic order features more interactions between the scalar and pseudoscalar glueballs determined by the following couplings:
\begin{equation}
    \begin{split}
\frac{\lambda_{12}}{\Lambda}&=2c_{2}\eta_{0}+3c_{4}a_{0}\,,\\
\frac{\lambda_{21}}{\Lambda}&=\frac{3\theta}{4}\eta_{0}+2c_{2}a_{0}\,,\\
\lambda_{22}&=c_{2}\,.\\
    \end{split}
\end{equation}
In particular, these interactions might allow for the decay of the pseudoscalar glueball, generally heavier than the scalar one. If the $a\to \eta\eta$ decay is kinematically allowed, $m_{a}>2m_{\eta}$, an important observable that can be calculated from Eq.~\eqref{eq:expapp} is the decay rate of the pseudoscalar glueball
\begin{equation}
    \Gamma_{a\to \eta\eta}=\frac{(\lambda_{21})^{2}\Lambda^{2}}{8\pi\,m_{a}}\sqrt{1-4\left(\frac{m_{\eta}}{m_{a}}\right)^{2}}\,.
    \label{eq:decapp}
\end{equation}
For a given decay rate, it is possible to set another constraint on the microscopic parameters of the theory. The system composed by Eqs.~\eqref{eq:V0app}-\eqref{eq:linapp} and \eqref{eq:decapp} is a system of non-linear equations that can be resolved in terms of seven unknown parameters such as $c_{i}$ for $i=0,\dots, 4$, $\eta_{0}$ and $a_{0}$. Its solution is neither unique nor guaranteed to exist for an arbitrary choice of the physical observables $m_{\eta}$, $m_{a}$ and $\Gamma_{a\to \eta\eta}$, but it would strongly reduce the overall number of free parameters of the considered EFT. This approach enables to determine, at least, some of the EFT parameters of Eq.~\eqref{eq:potential} as functions of the observable ones, whenever a proper solution exists.

\bibliographystyle{bibi}
\bibliography{biblio.bib}

\end{document}